\documentclass[12pt]{article}
\usepackage[utf8]{inputenc} 
\usepackage{cite}
\usepackage{graphicx}
\usepackage{times}
\usepackage{color}
\usepackage[labelfont=bf]{caption}

\newcommand*\eu{Eu$^{3+}$}
\newcommand{\fdz}{$^5$D$_{0}$}

\newcommand{\sfz}{$^7$F$_{0}$}

\newcommand*\trabs{$^7$F$_0\rightarrow ^5$D$_0$}

\newcommand*\PG{\textcolor{black}}

\title{Rare-Earth Molecular Crystals with Ultra-narrow Optical Linewidths for Photonic Quantum Technologies}

\date{\today}

\begin{document} 

\baselineskip24pt

\maketitle 
\begin{center}

\textbf {Diana Serrano$^1$, Kuppusamy Senthil Kumar$^{2,3}$, Beno\^{i}t Heinrich$^4$, Olaf Fuhr$^{3,5}$, David Hunger$^{2,6}$, Mario Ruben$^{2,3,7}$, Philippe Goldner$^1$.}\\
\vspace{1cm}
\normalsize{$^1$Chimie ParisTech, PSL University, CNRS, Institut de Recherche de Chimie Paris, 11, rue Pierre et Marie Curie, 75005 Paris, France}\\
\normalsize{$^{2}$Institute for Quantum Materials and Technologies (IQMT), Karlsruhe Institute of Technology (KIT), 76344 Eggenstein-Leopoldshafen, Germany}\\
\normalsize{$^{3}$ Institute of Nanotechnology, Karlsruhe Institute of Technology (KIT), 76344 Eggenstein-Leopoldshafen, Germany}\\
\normalsize{$^{4}${}Institut de Physique et Chimie des Mat\'eriaux de Strasbourg (IPCMS), CNRS-Universit\'e de Strasbourg, Strasbourg, France}\\
\normalsize{$^{5}$Karlsruhe Nano Micro Facility (KNMF), Karlsruhe Institute of Technology (KIT),}\\
\normalsize{$^{6}$Physikalisches Institut, Karlsruhe Institute of Technology (KIT), Wolfgang-Gaede Str. 1, 76131 Karlsruhe, Germany}\\
\normalsize{$^{7}$Centre Europ\'een de Sciences Quantiques (CESQ), Institut de Science et d'Ing\'enierie Supramol\'eculaire (ISIS), Universit\'e de Strasbourg, France}\\

\end{center}

\begin{abstract}
  Rare-earth ions are promising solid state systems to build light-matter interfaces at the quantum level. This relies on their potential to show narrow optical homogeneous linewidths or, equivalently, long-lived optical quantum states. In this letter, we report on europium molecular crystals that exhibit linewidths in the 10s of kHz range, orders of magnitude narrower than other molecular centers. We harness this property to demonstrate efficient optical spin initialization, coherent storage of light using an atomic frequency comb, and optical control of ion-ion interactions towards implementation of quantum gates. These results illustrate the utility of rare-earth molecular crystals as a new platform for photonic quantum technologies that combines highly coherent emitters with  the unmatched  versatility in composition, structure, and integration capability of molecular materials. 
\end{abstract}

Rare-earth ion (REI)-doped materials are promising
systems for optical quantum technologies. At cryogenic temperatures, REIs doped into high quality bulk single crystals, such as Y$_2$SiO$_5$, show exceptionally narrow  optical homogeneous linewidths, equivalent to long-lived quantum coherence \PG{lifetime }($T_2$) and suitable for building quantum light-matter interfaces \cite{goldner_rare_2015}. Moreover, REI \PG{can} present optically addressable electron and/or nuclear spin degrees of freedom that can be leveraged to efficiently store and process quantum information \cite{awschalom_quantum_2018,zhong_optically_2015}. These unique properties in the solid state have been used to demonstrate quantum memories for light 
\PG{\cite{de_riedmatten_solid-state_2008}}, light-matter teleportation \cite{bussieres_quantum_2014}, and frequency and time-multiplexed storage \cite{seri_quantum_2019}. REI-doped crystals are also actively investigated for optical to microwave conversion \cite{bartholomew_-chip_2020} and quantum processing \cite{kinos_roadmap_2021}. Besides experiments on high-quality bulk single crystals, strong efforts have recently been launched towards combining REIs with nanophotonic structures \cite{zhong_emerging_2019}. This has enabled single REI detection and control \cite{chen_parallel_2020}, \PG{fast} spontaneous emission modulation \cite{casabone_dynamic_2020}, lifetime limited single-photon emission \cite{zhong_optically_2018}, and on-chip optical storage \cite{zhong_nanophotonic_2017}. Further developments of these exciting topics is, however, impeded by the difficulty to nano-fabricate crystalline host materials 
\PG{that }preserve REIs quantum properties for integration into high-quality nanophotonic devices. Molecular chemistry is very attractive in this respect because it offers unmatched flexibility in terms of material composition, fine structural tuning, and integration into photonic structures, as demonstrated by numerous results obtained with single organic molecules embedded in crystalline host lattices \cite{toninelli_single_2020}. However, most organic molecules studied to date lack a spin degree of freedom. Recently, optically addressable molecular spins were demonstrated \cite{bayliss_optically_2020,kumar_optical_2021}, although with limited optical coherence, hindering their use as coherent spin-photon interfaces. Here, we introduce REI molecular crystals containing trivalent europium that exhibit optical homogeneous linewidths between 5 and 30 kHz, 3 to 4 orders of magnitude narrower than any molecular system \cite{kumar_optical_2021,zirkelbach_partial_2020,toninelli_single_2020}. This allows us to efficiently exploit the \eu\ nuclear spin degree of freedom by demonstrating $>$ 95\% spin \PG{initialization} 
into a single level, and coherent optical storage using atomic frequency combs. We also harness \eu\ narrow optical linewidth to demonstrate ion-ion interactions, which are the basis for high-bandwidth two-qubit quantum gates\cite{kinos_roadmap_2021}.

\begin{figure}
\includegraphics[width=\columnwidth]{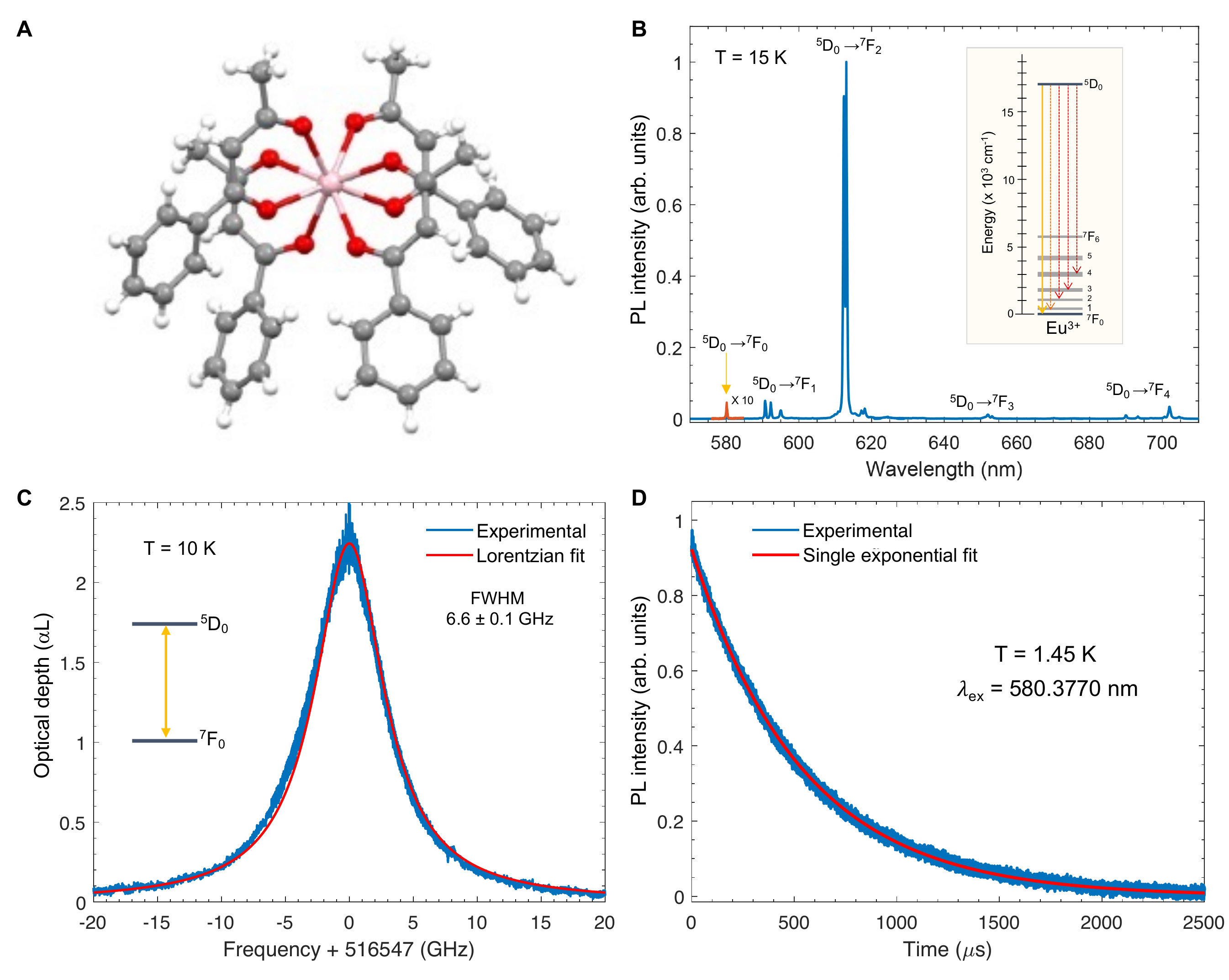}
\caption{\textbf{\PG{Material and} low temperature optical spectroscopy.} \textbf{A}. X-ray crystal structure of the \eu\ complex. The counter cation is omitted for clarity. Color code: grey, carbon; white, hydrogen; pink, europium; red, oxygen. \textbf{B}. \eu\ photo-luminescence (PL)  spectrum showing characteristic  $^5$D$_0 \to ^7$F$_J$ ($J=0-4$)  transitions  (see inset). All following results were obtained on the \trabs\ transition (\PG{in red}). \textbf{C}. $^7$F$_0 \rightarrow ^5$D$_0$ absorption line recorded on a 500 $\mu$m thick powder sample. Center wavelength: 580.3778 nm (vacuum). \textbf{D}. $^5$D$_0 \rightarrow ^7$F$_J$ ($J=1-4$) fluorescence decay. Red line: Single exponential fit to data giving a $^5$D$_0$ population lifetime of $T_{1,\mathrm{opt}}=540$ $\mu$s. \label{fig:panel1}}
\end{figure}

The molecular crystal is composed of a mononuclear \eu\ complex [Eu(BA)$_4$](pip), where BA and pip stand for benzoylacetonate and piperidin-1-ium, respectively (\textbf{Fig.} \ref{fig:panel1}\textbf{A}) \cite{melby_synthesis_1964}. For clarity, it is referred to as \eu\ complex hereafter. The complex crystallized in P2$_1$/n space group, belonging to the monoclinic crystal system, with crystal lattice composed of anionic [Eu(BA)$_4$]$^-$ and cationic piperidin-1-ium units as shown in the supplementary information. The all-oxygen coordination environment around the \eu\ center is best described as a biaugmented trigonal prism with the help of a continuous shape measure calculation (CShM) \cite{binnemans_interpretation_2015}. The point group symmetry around the \eu\ center is assigned to C$_{2v}$, as inferred from the CShM calculation and luminescence spectrum depicted in \textbf{Fig.} \ref{fig:panel1}\textbf{B}. All experiments discussed in the following were performed on crystalline powders with grain sizes of at least 50 nm. The \eu\ complex showed high physical and chemical stabilities over time. Repeated cooling cycles had no noticeable effects on optical properties, and no photo-degradation was observed neither at low temperature nor under high laser intensity.

\textbf{Fig.} \ref{fig:panel1}B shows the emission spectrum of the \eu\ complex where the characteristic lines of trivalent europium are observed \cite{binnemans_interpretation_2015}. We focused on the \trabs\ transition at 580.3778 nm (vacuum) since this transition is associated with narrow linewidths in crystals like Y$_2$SiO$_5$ \cite{konz_temperature_2003}. Moreover, it enables optical control of ground state nuclear spins, an important feature for \PG{applications in} quantum technologies. Transmission experiments revealed a  \trabs\ inhomogeneous linewidth ($\Gamma_\mathrm{inh}$) of 6.6 GHz, corresponding to 0.007 nm (\textbf{Fig.} \ref{fig:panel1}\textbf{C}). This low value, together with a Lorentzian absorption profile, is typical of high crystalline quality samples with low disorder, and comparable to values obtained in crystals like \eu :Y$_2$O$_3$ \cite{zhong_emerging_2019}. Strong absorption was observed in a 500 $\mu$m thick powder as a result of the stoichiometric composition of the molecular crystal and light scattering \PG{inside the powder}. Fluorescence decay experiments led to a lifetime value $T_{1,\mathrm{opt}}$ of 540 $\mu$s for the $^5$D$_0$ level (\textbf{Fig.} \ref{fig:panel1}\textbf{D}). This sets a limit of only $1/2\pi T_{1,\mathrm{opt}} = 295$ Hz on the optical homogeneous linewidth ($\Gamma_h$).

\begin{figure}
     \centering
     \includegraphics[width=\columnwidth]{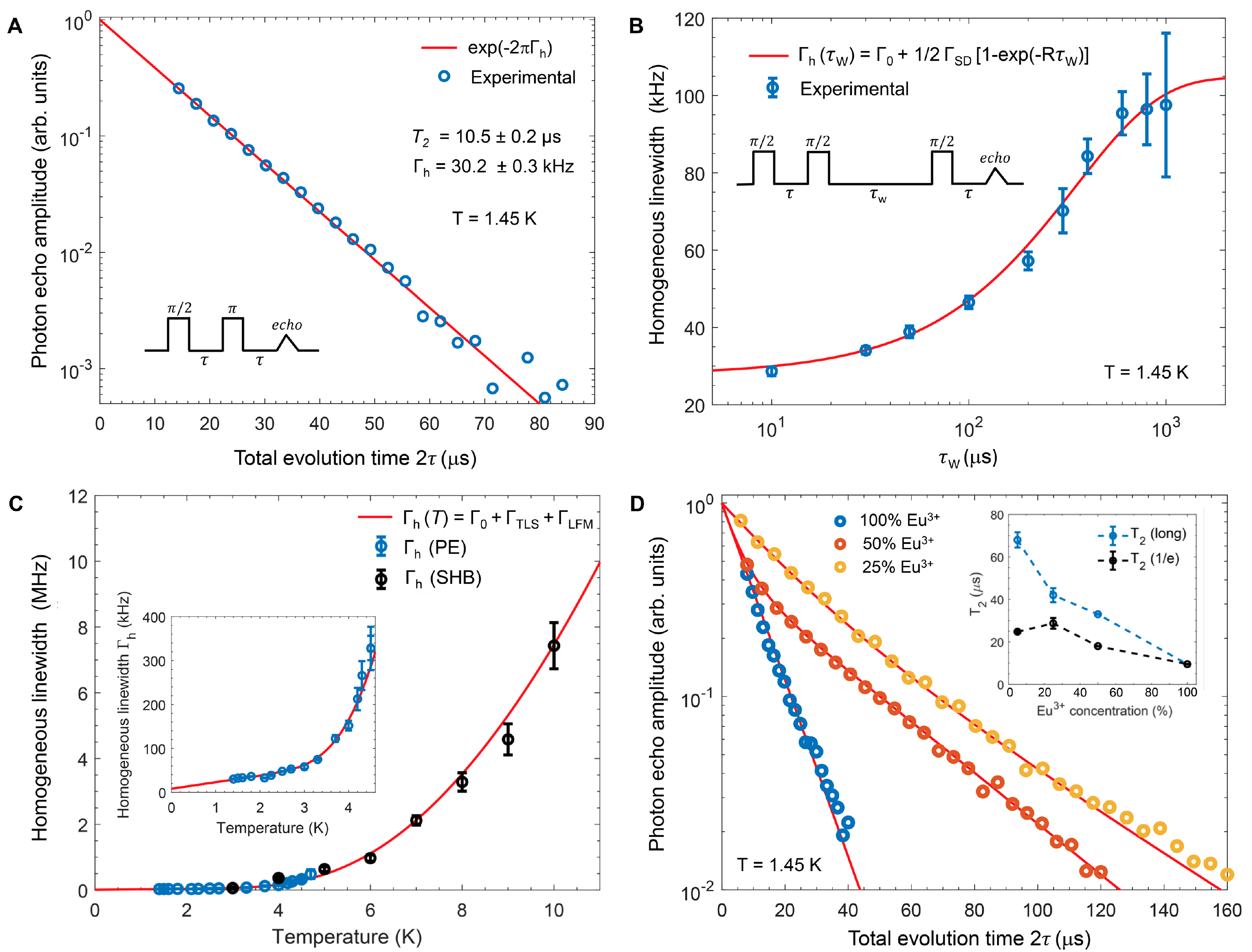}
      \caption{\textbf{Ultra-narrow optical homogeneous linewidths.}
     \textbf{A}. 2-pulse photon echo (PE, \PG{pulse sequence shown in inset}) decay for the $^7$F$_0 \rightarrow ^5$D$_0$ transition. Red line: exponential fit to data giving $T_2 = 10.5 \pm 0.2$ $\mu$s or $\Gamma_h = 1/\pi T_2 = 30.2 \pm 0.2$ kHz. \textbf{B}. $\Gamma_h$ as a function of waiting time ($\tau_W$) measured by 3-pulse PE. Red line: fit using a spectral diffusion model yielding a flip rate $R=2.9 \pm 0.8$ kHz and a width $\Gamma_{SD} = 154 \pm 16$ kHz.  \textbf{C}. Temperature dependence of $\Gamma_h$ measured by 2-pulse PE (blue dots) and spectral hole burning (black dots). Red line: fit to a model including contributions from two-level systems (TLS) and quasi-localized low frequency modes (LFMs). Inset: Zoom on lower temperature range. \textbf{D}. PE decays from Y$^{3+}$-diluted complexes showing $T_2$ increase with decreasing \eu\ concentration. Red line: single (non diluted crystal) and bi- (diluted crystals) exponential decay fits to data. Inset: 1/e echo amplitude decay time $T_2$ (black dots) and long $T_2$ component from bi-exponential fits (blue dots) as a function of \eu\ concentration.}
     \label{fig:2}
 \end{figure}
The photon echo (PE) technique, similar to Hahn's spin echo, enables accurate assessment of narrow homogeneous linewidths\cite{abella_photon_1966,goldner_rare_2015}. Due to its high sensitivity, it can be applied to powders \cite{perrot_narrow_2013}, allowing us to measure the optical coherence lifetime, or quantum state lifetime $T_2$, of the \fdz$\leftrightarrow$\sfz\ transition in the \eu\ complex (\textbf{Fig.} \ref{fig:2}\textbf{A}). $\Gamma_h=1/\pi T_2$  was found equal to 30.2 $\pm$ 0.2 kHz, remarkably lower than values reported for single molecules \cite{toninelli_single_2020}, transition metal ions \cite{riesen_hole-burning_2006}, and a previously reported \eu\ complex \cite{kumar_optical_2021}, all  in the 10s of MHz range. The \eu\ linewidth in the complex is indeed comparable to those measured in some bulk crystals, such as EuP$_5$O$_{14}$ \cite{shelby_frequency-dependent_1980}, and in \eu:Y$_2$O$_3$ nanoparticles \cite{zhong_emerging_2019}. We further explored line broadening using stimulated photon echoes (or 3-pulse PE) which allows measuring $\Gamma_h$ over the time scale of the excited state population lifetime $T_{1,\mathrm{opt}}$. We observed an increase in homogeneous linewidth with increasing waiting time $\tau_W$ (\textbf{Fig.} \ref{fig:2}\textbf{B}) until about $1$ ms where $\Gamma_h$ plateaued at a value of $\approx 105$  kHz. The $\Gamma_h$ evolution has an s-shape, as already observed in other REI-doped crystals, and was modelled using a sudden-jump spectral diffusion (SD) model, which could indicate interactions with defects or impurities carrying electron spins \cite{bottger_optical_2006}. The limited line broadening over 100s of $\mu$s in the \eu\ complex is favorable for spectrally resolved repetitive single ion addressing. Furthermore, lifetime shortening by Purcell enhancement in an optical micro-cavity by a factor of $\approx 100$, would enable generation of indistinguishable single photons  \cite{merkel_coherent_2020, casabone_dynamic_2020}.

Insights into dephasing mechanisms occurring in the complex were also obtained from the evolution of $\Gamma_h$ with temperature $T$, measured by 2-pulse PE and spectral hole burning (SHB) (\textbf{Fig.} \ref{fig:2}\textbf{C}). Two regimes were identified: for temperatures below 3.5 K, dephasing is dominated by coupling to two-level systems (TLS) and $\Gamma_h$ increases linearly with $T$  (\textbf{Fig.} \ref{fig:2}\textbf{C} inset) \cite{flinn_sample-dependent_1994}; above 3.5 K, the exponential  increase of $\Gamma_h$ is attributed to quasi-localized low frequency modes (LFMs) \cite{kozankiewicz_single-molecule_2014}.  Data were modeled using the expression $\Gamma_{h}(T) = \Gamma_{0} + \Gamma_\mathrm{TLS} + \Gamma_\mathrm{LFM}$, where $\Gamma_\mathrm{TLS} = \alpha_\mathrm{TLS}T$ and $\Gamma_{LFM} \approx \alpha_\mathrm{LFM} \exp(-\Delta E_\mathrm{LFM}/k_BT)$, with $k_B$ the Boltzmann constant. The best fit gave  a TLS rate $\alpha_\mathrm{TLS}  = 15 \pm 5$ kHz K$^{-1}$, as observed in some REI-doped crystals \cite{flinn_sample-dependent_1994}, and LFM transition energy $\Delta E_\mathrm{LFM} = 580$ GHz (19 cm$^{-1}$), in the range reported for single molecules \cite{kozankiewicz_single-molecule_2014}. The homogeneous linewidth extrapolated to 0 K ($\Gamma_{0}$) is estimated at 8 $\pm$ 4 kHz. This remaining dephasing could be due to \eu-\eu\ interactions\cite{konz_temperature_2003}, as the crystal has a high \eu\ concentration, $C_\mathrm{Eu} = 9.6 \times 10^{20}$ ions cm$^{-3}$. To investigate this effect, an optically inactive REI, Y$^{3+}$, was introduced in the crystal to reduce $C_\mathrm{Eu}$. \textbf{Fig.} \ref{fig:2}\textbf{D} shows 2-pulse PE decays in a series of diluted crystals. The non-exponential decays suggest distinct populations or environments for the \eu\ ions in these complexes. Increasing coherence lifetimes with decreasing  $C_\mathrm{Eu}$ are observed, especially for the long decay components (\textbf{Fig.} \ref{fig:2} \textbf{D}, inset), confirming a contribution of \eu-\eu\ interactions to dephasing. At the highest dilution investigated here (95\% that is 5\% \eu\ content), we found a $1/e$ decay time  $T_{2} =  25$ $\mu$s, that is $\Gamma_h = 12.7$ kHz. Remarkably, in this sample, photon echoes could be detected even after evolution times of 300 $\mu$s, corresponding to a long decay component of $T_2 = 68 \pm 4$ $\mu$s ($\Gamma_h= 4.6 \pm 0.2$ kHz) (\textbf{Fig.} \ref{fig:2}\textbf{D} inset). This suggests that synthesis could be optimized to provide a nearly noise-free environment to all \eu\ ions in diluted samples.  

\begin{figure}
     \centering
     \includegraphics[width=\columnwidth]{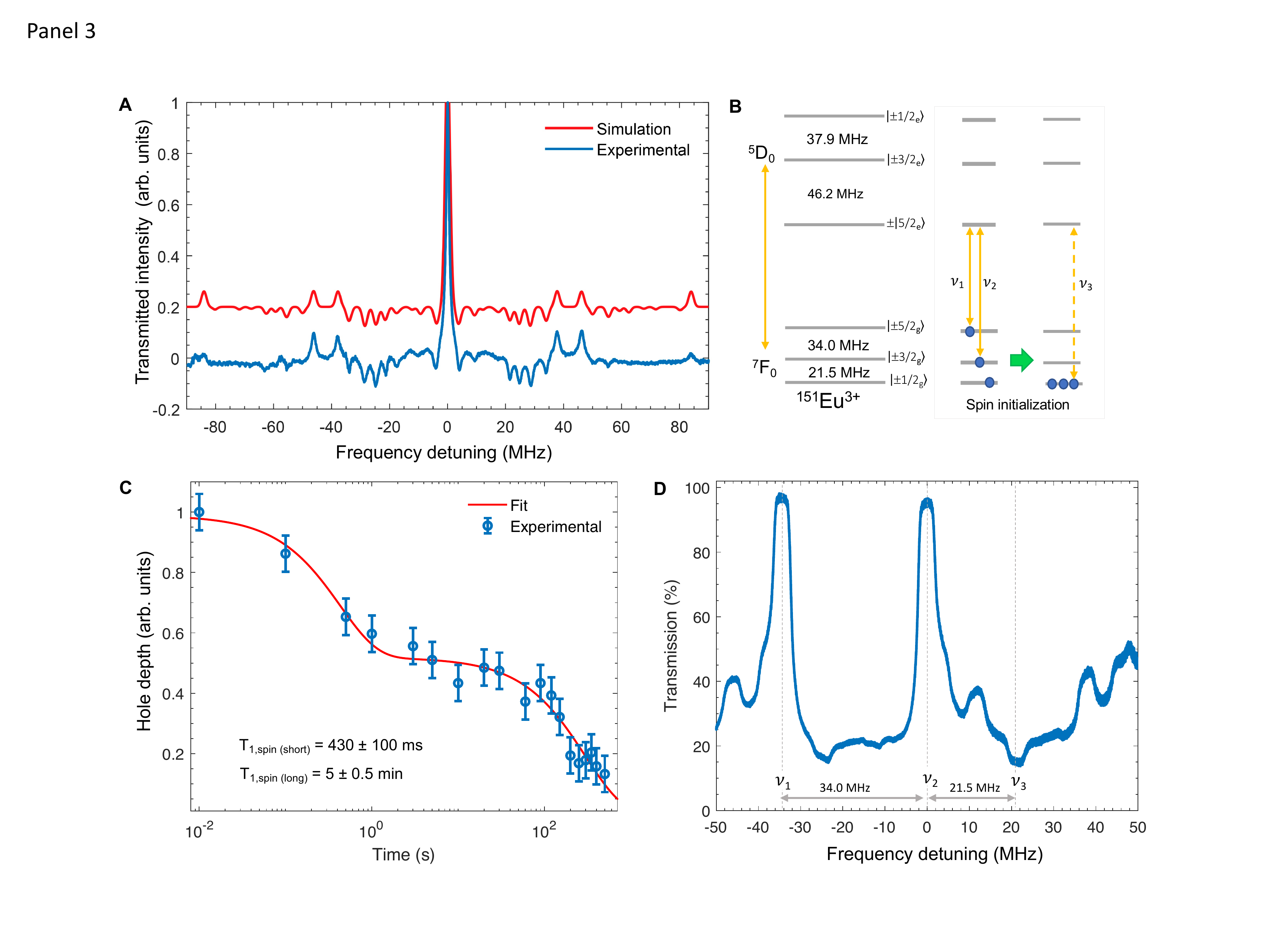}
     \caption{\textbf{Optically addressable nuclear spins.} \textbf{A}. Spectral hole burning experimental spectrum of the $^{151}$\eu\ complex (blue) compared to simulation (red). \textbf{B}. Left -  $^{151}$\eu\ ground and excited state nuclear spin splittings. Right - Scheme of spin initialisation by two-color optical pumping. \textbf{C}. Hole decay as a function of delay between burn and readout pulses (see Methods in SI). Red line: double exponential fit yielding spin relaxation times $ T_{1,\mathrm{spin}}$ of $430 \pm 100$ ms and $5.0 \pm 0.5$ min. \textbf{D}. Initialisation to a single spin level by \PG{two-color} optical pumping over a bandwidth of 3 MHz. Applying the pumping scheme shown in B (right), population is transferred from $|\pm3/2_g>$ and $|\pm5/2_g>$ levels (deep holes at $\nu_1$ and $\nu_2$), to $|\pm1/2_g>$ (increased absorption, or anti-hole, at $\nu_3$). \PG{Center wavelength 580.3778 nm, $T = 1.45$ K.}}
     \label{fig:3}
 \end{figure}

Narrow optical linewidths are key to efficiently address REI spins, which enabled detailed investigations of \eu\ nuclear spin states in the complex. 

First, the nuclear spin structure was determined for the \fdz\ ground state and \sfz\ excited states by SHB in a complex isotopically enriched in $^{151}$\eu{}. At zero magnetic field, it consists of three doubly-degenerate levels split by quadrupolar interaction \cite{konz_temperature_2003}, giving rise to a complex SHB spectrum after optical pumping. Thanks to the narrow optical linewidth, well-resolved spectral features could be recorded and analyzed (\textbf{Fig.} \ref{fig:3}\textbf{A}). This allowed us to assign ground and excited state splitting energies (\textbf{Fig.} \ref{fig:3}\textbf{B}-Left), and transition branching ratios between nuclear spin levels. The same study was done in a natural abundance sample, from which, splitting energies and transition branching ratios were determined for the $^{153}$\eu\ isotope. Importantly, the obtained branching ratios show the existence of efficient three-level lambda systems for both isotopes in the \eu\ complex, an important requirement for all-optical spin control \cite{serrano_all-optical_2018}.

We then probed the dynamics of ground state spin populations by monitoring the spectral hole depth as a function of time delay between burn and readout pulses (\textbf{Fig.} \ref{fig:3}\textbf{C}). Two distinct decay components can be observed with relaxation times estimated at $430 \pm 100$ ms and $5.0 \pm 0.5$ min. This shows that nuclear spin levels can be used as shelving states for at least  100s of ms. Such long times enable efficient optical manipulation of the spin population. As an example, we initialised ions into one nuclear spin level using two-color laser pulses at frequencies $\nu_1$ and $\nu_2$. This simultaneously depletes the $|\pm 5/2_g>$ and $|\pm 3/2_g>$ ground state spin levels and transfers population into the $|\pm 1/2_g>$ (\textbf{Fig.} \ref{fig:3}\textbf{B}-Right. As displayed in \textbf{Fig.} \ref{fig:3}\textbf{D}, nearly full transparency could be induced at $\nu_1$ and $\nu_2$, which translates to $> 95$\% spin population into a single level. This also proves that efficient spectral tailoring is possible in the \eu\ complex, an essential feature for many quantum memory and processing protocols based on REIs\cite{de_riedmatten_solid-state_2008,kinos_roadmap_2021}.

We next used the narrow optical linewidths of \eu\ molecular crystals to demonstrate coherent optical storage and controlled ion-ion interactions. In the first case, we used the atomic frequency comb (AFC) protocol \cite{de_riedmatten_solid-state_2008} to store a light pulse in the molecular crystal. 

\begin{figure}
\includegraphics[width=\columnwidth]{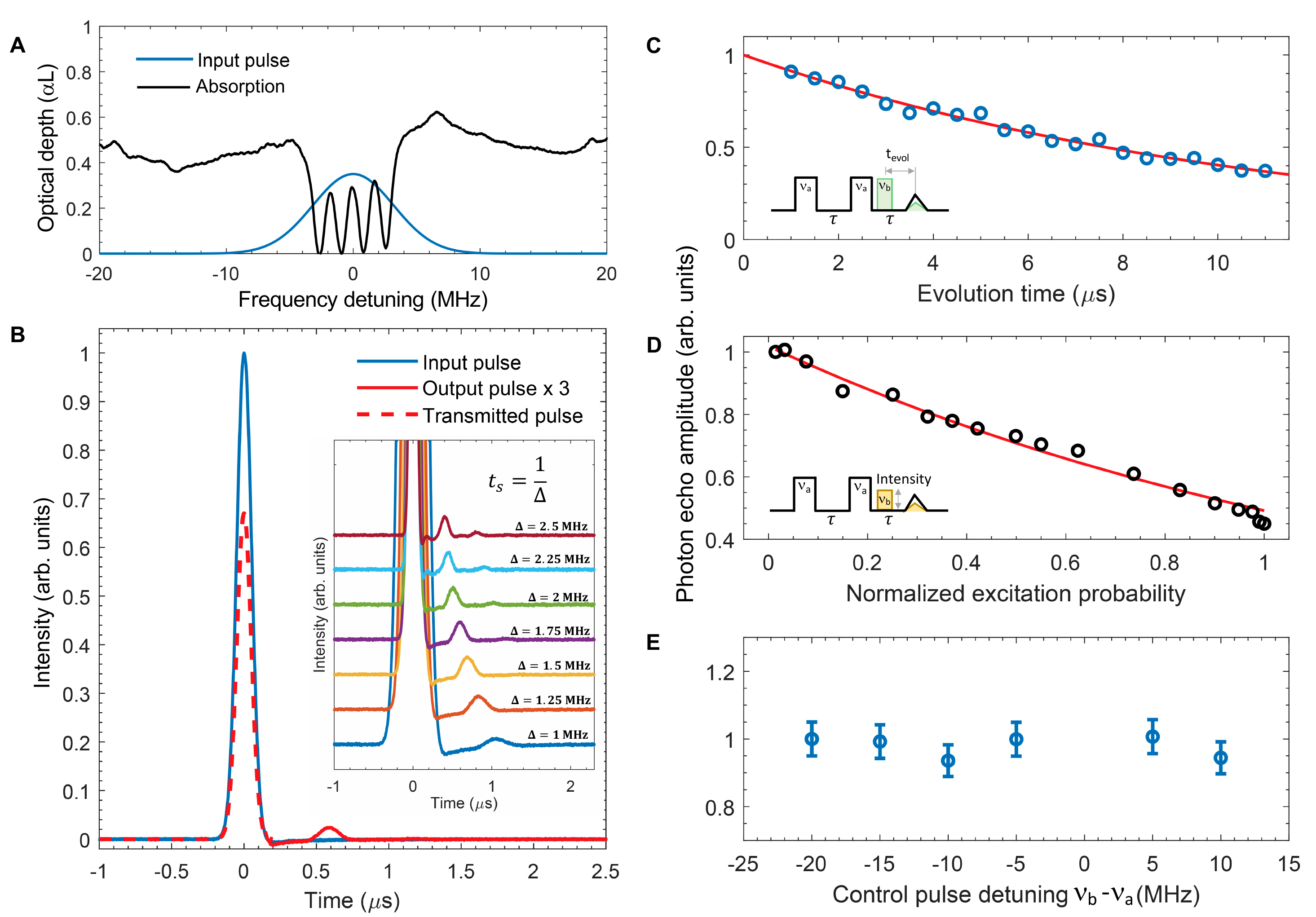}
\caption{\textbf{Coherent light storage and optically controlled ion-ion interactions.} \textbf{A}. Atomic frequency comb (AFC)  created by optical pumping within the inhomogeneously broadened \trabs\ line. The 0.9 MHz wide teeth are separated by 1.75 MHz and span 6 MHz. Blue line: \PG{Input pulse spectrum. Center wavelength 580.373 nm}.\textbf{B}. Coherent storage of a 0.15 $\mu$s long input pulse (blue line) using the AFC structure in \textbf{A}. The output pulse appears at $0.57 \mu$s (solid red line, intensity $\times$ 3) with a storage efficiency of  0.86\% (see text). Part of the input pulse is directly transmitted and not stored (dashed line at 0 $\mu$s). Inset: variable storage times \PG{$t_\mathrm{s}$} obtained by changing AFC teeth frequency separation \PG{$\Delta$}. Input pulse \PG{lengths were adapted to keep identical spectral overlap with AFCs.}  \textbf{C,D,E}. Photon echo amplitude of target ions (at frequency $\nu_a$) as a function of: (\textbf{C}) evolution time $t_\mathrm{evol}$ after excitation of control ions \PG{(at $\nu_b = \nu_a-20$ MHz)}. Red line: exponential fit to data giving a broadening of the target ions linewidth by the control ions of $\Gamma_c = 14.5 $ kHz (see text); (\textbf{D}) control ions excitation probability \PG{(normalized to the maximum value reached experimentally)}. Red line: exponential fit; (\textbf{E}) control pulse detuning. All experiments were performed at 1.45 K.\label{fig:panel4}} 
\end{figure}

This protocol enables very low output noise and multiplexed storage, important assets for long distance quantum communications. We first made use of the efficient optical pumping achieved in the \eu\ complex to create an AFC spanning a 6 MHz range, with three 0.9 MHz wide teeth separated by 1.75 MHz (\textbf{Fig.} \ref{fig:panel4}\textbf{A}). A storage experiment was then performed with a 0.15 $\mu$s long input pulse overlapping the AFC structure. As shown in \textbf{Fig.} \ref{fig:panel4}\textbf{B}, the output pulse is observed at a delay $t_s =0.57$ $\mu$s after the partially transmitted input pulse, in perfect agreement with the teeth spacing since $1/t_s = 1.75$ MHz \cite{de_riedmatten_solid-state_2008}. By varying the teeth spacing and adapting the input pulse length to keep an identical spectral overlap with the comb, output pulses were observed up to about 1 $\mu$s storage time, clearly confirming the AFC  process (\textbf{Fig.} \ref{fig:panel4}\textbf{B} inset). Storage efficiency,  defined as the ratio between input and output pulses intensities, was 0.86\% for a storage time $t_s = 0.57$ $\mu$s, in agreement with theory. This efficiency could be boosted up to 100\% in an optical cavity \cite{afzelius_impedance-matched_2010}. This could be achieved by crystallizing the \eu\ complex directly on a cavity mirror. Storage times up to  several 10s of $\mu$s could also be obtained by creating narrower teeth \cite{de_riedmatten_solid-state_2008} with a suitable laser, given the narrow homogeneous linewidths and limited spectral diffusion in the \eu\ complex.

We finally investigated controlled interactions between \eu\ ions. The scheme we used is based on the difference in permanent electric dipole moments between \eu\ ground (\sfz) and excited (\fdz) states \cite{macfarlane_optical_2007}. This difference occurs when \eu\ ions sit in a low symmetry site, \PG{which is the case in our molecular crystal ($C_{2v}$ site symmetry).}
When a control ion is excited, the electric field it produces changes, causing a shift in transition frequency for a nearby target \eu\ ion through the linear Stark effect. This mechanism is the basis for 2-qubit gates and qubit readout in some REI-based quantum computing proposals \cite{kinos_roadmap_2021}. However, to be useful, ion-ion interactions must be significantly larger than the optical homogeneous linewidth, 
a condition well matched by our highly-concentrated \eu complex with 10 kHz linewidth. 

Target and control ions were chosen at different frequencies within the absorption line (\textbf{Fig.} \ref{fig:panel1}\textbf{C}) to allow for independent excitation and monitoring. Because of the distribution of distances and orientations and therefore interaction strengths between \eu\ ions, the excitation of control ions results in an additional line broadening $\Gamma_c$ for target ions \cite{altner_dephasing-rephasing_1996}. $\Gamma_c$ can be conveniently  measured by monitoring the amplitude of a photon echo produced by target ions while an extra pulse excites the control ions. 

We first investigated the effect of changing the evolution time ($t_\mathrm{evol}$) between control pulse and echo, as shown in \textbf{Fig.} \ref{fig:panel4}\textbf{C}. In this case, the echo amplitude varies as 
$\exp (-\pi \Gamma_c t_\mathrm{evol})$ \cite{altner_dephasing-rephasing_1996}.
A fit to the experimental data yields $\Gamma_c = 14.5$ kHz, in qualitative agreement with expected \eu$-$\eu\ electric dipole interactions and previous experiments in non-molecular REI-doped crystals \cite{altner_dephasing-rephasing_1996}. It also provides an upper bound to the ion-ion interaction contribution in the measurements of $\Gamma_h$ by 2-pulse photon echoes. We further confirmed this analysis by varying the control pulse intensity $I_c$. In the weak excitation regime, $\Gamma_c$ is proportional to 
the fraction of excited control ions ($p$), with $p \propto I_c$.  We indeed observed the predicted exponential decay of the echo amplitude with increasing $p$, as displayed in Fig.  \ref{fig:panel4}\textbf{D}. Finally, the control pulse frequency was varied over several 10s of MHz with fixed $t_\mathrm{evol}$ and $p$. No significant change in echo amplitude was observed, ruling out direct light-induced frequency shift of target ions, the so called AC Stark shift (\textbf{Fig.} \ref{fig:panel4}\textbf{E}) \cite{zhong_nanophotonic_2017}. We therefore conclude that qubit gate and readout schemes based on electric dipole-dipole interactions could be implemented in the molecular crystal. Because of the high \eu\ concentration and narrow linewidth, we estimate that one ion could control  thousands of  target ions, a useful property for scaling up REI-based quantum processors \cite{kinos_roadmap_2021}. 

The \eu\ complex investigated in this study is a very promising platform for optical quantum technologies as a robust system with narrow linewidth and long-lived optically addressable spins.  Furthermore, significant decrease in optical homogeneous linewidth could be obtained by lowering temperature, tuning the REI concentration, and optimizing synthesis to decrease residual defects or impurities.  
In such compounds, interactions with host nuclear spins like $^1$H or $^{13}C$  could become  the dominant dephasing mechanisms, which can be reduced using specific isotopes with lower or zero magnetic moments \cite{wernsdorfer_synthetic_2019}. This could also be useful to obtain spin states with long coherence lifetimes, a property that remains to be measured in our samples. The huge possibility in molecular design also opens the way to accurate engineering of the ligand field acting on europium and more generally other REIs of interest for quantum technologies, such as Er$^{3+}$ or Yb$^{3+}$. This would allow one to optimize transition strengths and frequencies for optimal coupling to light, tune electronic and spin level structures, and coupling to molecular vibrations for long coherence lifetimes. Complexes containing several REIs at close distance can also be synthesized which would enable high-density optically addressable qubit architectures, as shown with REI magnetic molecules in the microwave range \cite{godfrin_operating_2017}.  
Developments towards applications in quantum technologies could greatly benefit from integration of molecular crystals into nanophotonic structures. For example, high-quality and small mode volume optical cavities can dramatically enhance coupling of REI  with light, enabling efficient emission from nominally weak transitions\cite{kindem_control_2020,chen_parallel_2020,casabone_dynamic_2020}. 
This would be facilitated by the broad range of synthesis, functionalization and deposition methods that are available for molecular complexes \cite{wernsdorfer_synthetic_2019,toninelli_single_2020}, as well as the scalable production of large amounts of identical molecules. REI molecular crystals could  therefore emerge as a highly versatile platform for highly coherent  light-matter-spin quantum interfaces for developing applications in quantum communications and processing and fundamental studies in quantum optics.

\bibliographystyle{ieeetr}
\bibliography{references}

\begin{thebibliography}{10}

\bibitem{goldner_rare_2015}
P.~Goldner, A.~Ferrier, and O.~Guillot-Noël, ``Rare {Earth}-{Doped} {Crystals}
  for {Quantum} {Information} {Processing},'' in {\em Handbook on the {Physics}
  and {Chemistry} of {Rare} {Earths}} (J.-C.~G. Bünzli and V.~K. Pecharsky,
  eds.), vol.~46, pp.~1--78, Amsterdam: Elsevier, 2015.

\bibitem{awschalom_quantum_2018}
D.~D. Awschalom, R.~Hanson, J.~Wrachtrup, and B.~B. Zhou, ``Quantum
  technologies with optically interfaced solid-state spins,'' {\em Nature
  Photonics}, vol.~12, pp.~516--527, Aug. 2018.

\bibitem{zhong_optically_2015}
M.~Zhong, M.~P. Hedges, R.~L. Ahlefeldt, J.~G. Bartholomew, S.~E. Beavan, S.~M.
  Wittig, J.~J. Longdell, and M.~J. Sellars, ``Optically addressable nuclear
  spins in a solid with a six-hour coherence time,'' {\em Nature}, vol.~517,
  no.~7533, pp.~177--180, 2015.

\bibitem{de_riedmatten_solid-state_2008}
H.~de~Riedmatten, M.~Afzelius, M.~U. Staudt, C.~Simon, and N.~Gisin, ``A
  solid-state light–matter interface at the single-photon level,'' {\em
  Nature}, vol.~456, no.~7223, pp.~773--777, 2008.

\bibitem{bussieres_quantum_2014}
F.~Bussières, C.~Clausen, A.~Tiranov, B.~Korzh, V.~B. Verma, S.~W. Nam,
  F.~Marsili, A.~Ferrier, P.~Goldner, H.~Herrmann, C.~Silberhorn, W.~Sohler,
  M.~Afzelius, and N.~Gisin, ``Quantum teleportation from a telecom-wavelength
  photon to a solid-state quantum memory,'' {\em Nature Photonics}, vol.~8,
  pp.~775--778, 2014.

\bibitem{seri_quantum_2019}
A.~Seri, D.~Lago-Rivera, A.~Lenhard, G.~Corrielli, R.~Osellame, M.~Mazzera, and
  H.~de~Riedmatten, ``Quantum {Storage} of {Frequency}-{Multiplexed} {Heralded}
  {Single} {Photons},'' {\em Physical Review Letters}, vol.~123, p.~080502,
  Aug. 2019.

\bibitem{bartholomew_-chip_2020}
J.~G. Bartholomew, J.~Rochman, T.~Xie, J.~M. Kindem, A.~Ruskuc, I.~Craiciu,
  M.~Lei, and A.~Faraon, ``On-chip coherent microwave-to-optical transduction
  mediated by ytterbium in {YVO4},'' {\em Nature Communications}, vol.~11,
  p.~3266, Dec. 2020.

\bibitem{kinos_roadmap_2021}
A.~Kinos, D.~Hunger, R.~Kolesov, K.~Mølmer, H.~de~Riedmatten, P.~Goldner,
  A.~Tallaire, L.~Morvan, P.~Berger, S.~Welinski, K.~Karrai, L.~Rippe,
  S.~Kröll, and A.~Walther, ``Roadmap for {Rare}-earth {Quantum}
  {Computing},'' {\em arXiv:2103.15743 [quant-ph]}, Mar. 2021.

\bibitem{zhong_emerging_2019}
T.~Zhong and P.~Goldner, ``Emerging rare-earth doped material platforms for
  quantum nanophotonics,'' {\em Nanophotonics}, vol.~8, pp.~2003--2015, Nov.
  2019.

\bibitem{chen_parallel_2020}
S.~Chen, M.~Raha, C.~M. Phenicie, S.~Ourari, and J.~D. Thompson, ``Parallel
  single-shot measurement and coherent control of solid-state spins below the
  diffraction limit,'' {\em Science}, vol.~370, pp.~592--595, Oct. 2020.

\bibitem{casabone_dynamic_2020}
B.~Casabone, C.~Deshmukh, S.~Liu, D.~Serrano, A.~Ferrier, T.~Hümmer,
  P.~Goldner, D.~Hunger, and H.~de~Riedmatten, ``Dynamic control of {Purcell}
  enhanced emission of erbium ions in nanoparticles,'' {\em arXiv:2001.08532},
  vol.~quant-ph, Jan. 2020.

\bibitem{zhong_optically_2018}
T.~Zhong, J.~M. Kindem, J.~G. Bartholomew, J.~Rochman, I.~Craiciu, V.~Verma,
  S.~W. Nam, F.~Marsili, M.~D. Shaw, A.~D. Beyer, and A.~Faraon, ``Optically
  {Addressing} {Single} {Rare}-{Earth} {Ions} in a {Nanophotonic} {Cavity},''
  {\em Physical Review Letters}, vol.~121, p.~183603, Oct. 2018.

\bibitem{zhong_nanophotonic_2017}
T.~Zhong, J.~M. Kindem, J.~G. Bartholomew, J.~Rochman, I.~Craiciu, E.~Miyazono,
  M.~Bettinelli, E.~Cavalli, V.~Verma, S.~W. Nam, F.~Marsili, M.~D. Shaw, A.~D.
  Beyer, and A.~Faraon, ``Nanophotonic rare-earth quantum memory with optically
  controlled retrieval,'' {\em Science}, vol.~357, pp.~1392--1395, Sept. 2017.

\bibitem{toninelli_single_2020}
C.~Toninelli, I.~Gerhardt, A.~S. Clark, A.~Reserbat-Plantey, S.~Götzinger,
  Z.~Ristanovic, M.~Colautti, P.~Lombardi, K.~D. Major, I.~Deperasińska, W.~H.
  Pernice, F.~H.~L. Koppens, B.~Kozankiewicz, A.~Gourdon, V.~Sandoghdar, and
  M.~Orrit, ``Single organic molecules for photonic quantum technologies,''
  {\em arXiv:2011.05059 [quant-ph]}, Nov. 2020.

\bibitem{bayliss_optically_2020}
S.~L. Bayliss, D.~W. Laorenza, P.~J. Mintun, B.~D. Kovos, D.~E. Freedman, and
  D.~D. Awschalom, ``Optically addressable molecular spins for quantum
  information processing,'' {\em Science}, vol.~370, pp.~1309--1312, Dec. 2020.

\bibitem{kumar_optical_2021}
K.~S. Kumar, D.~Serrano, A.~M. Nonat, B.~Heinrich, L.~Karmazin, L.~J.
  Charbonnière, P.~Goldner, and M.~Ruben, ``Optical spin-state polarization in
  a binuclear europium complex towards molecule-based coherent light-spin
  interfaces,'' {\em Nature Communications}, vol.~12, p.~2152, Dec. 2021.

\bibitem{zirkelbach_partial_2020}
J.~Zirkelbach, B.~Gmeiner, J.~Renger, P.~Türschmann, T.~Utikal, S.~Götzinger,
  and V.~Sandoghdar, ``Partial {Cloaking} of a {Gold} {Particle} by a {Single}
  {Molecule},'' {\em Physical Review Letters}, vol.~125, p.~103603, Sept. 2020.

\bibitem{melby_synthesis_1964}
L.~R. Melby, N.~J. Rose, E.~Abramson, and J.~C. Caris, ``Synthesis and
  {Fluorescence} of {Some} {Trivalent} {Lanthanide} {Complexes},'' {\em Journal
  of the American Chemical Society}, vol.~86, pp.~5117--5125, Dec. 1964.

\bibitem{binnemans_interpretation_2015}
K.~Binnemans, ``Interpretation of europium ({III}) spectra,'' {\em Coordination
  Chemistry Reviews}, vol.~295, pp.~1--45, 2015.

\bibitem{konz_temperature_2003}
F.~Könz, Y.~Sun, C.~W. Thiel, R.~L. Cone, R.~Equall, R.~Hutcheson, and R.~M.
  Macfarlane, ``Temperature and concentration dependence of optical dephasing,
  spectral-hole lifetime, and anisotropic absorption in {Eu3}+:{Y2SiO5},'' {\em
  Physical Review B}, vol.~68, no.~8, p.~085109, 2003.

\bibitem{abella_photon_1966}
I.~D. Abella, N.~A. Kurnit, and S.~R. Hartmann, ``Photon echoes,'' {\em
  Physical Review}, vol.~141, no.~1, p.~391, 1966.

\bibitem{perrot_narrow_2013}
A.~Perrot, P.~Goldner, D.~Giaume, M.~Lovrić, C.~Andriamiadamanana, R.~R.
  Gonçalves, and A.~Ferrier, ``Narrow {Optical} {Homogeneous} {Linewidths} in
  {Rare} {Earth} {Doped} {Nanocrystals},'' {\em Physical Review Letters},
  vol.~111, no.~20, p.~203601, 2013.

\bibitem{riesen_hole-burning_2006}
H.~Riesen, ``Hole-burning spectroscopy of coordination compounds,'' {\em
  Coordination Chemistry Reviews}, vol.~250, pp.~1737--1754, July 2006.

\bibitem{shelby_frequency-dependent_1980}
R.~Shelby and R.~M. Macfarlane, ``Frequency-{Dependent} {Optical} {Dephasing}
  in the {Stoichiometric} {Material} {EuP}\_\{5\}{O}\_\{14\},'' {\em Physical
  Review Letters}, vol.~45, pp.~1098--1101, Sept. 1980.

\bibitem{bottger_optical_2006}
T.~Böttger, C.~W. Thiel, Y.~Sun, and R.~L. Cone, ``Optical decoherence and
  spectral diffusion at 1.5 $\mu$m in {Er3}+:{Y2SiO5} versus magnetic field,
  temperature, and {Er3}+ concentration,'' {\em Physical Review B}, vol.~73,
  no.~7, p.~075101, 2006.

\bibitem{merkel_coherent_2020}
B.~Merkel, A.~Ulanowski, and A.~Reiserer, ``Coherent and {Purcell}-{Enhanced}
  {Emission} from {Erbium} {Dopants} in a {Cryogenic} {High}- {Q}
  {Resonator},'' {\em Physical Review X}, vol.~10, p.~041025, Nov. 2020.

\bibitem{flinn_sample-dependent_1994}
G.~P. Flinn, K.~W. Jang, J.~Ganem, M.~L. Jones, R.~S. Meltzer, and R.~M.
  Macfarlane, ``Sample-dependent optical dephasing in bulk crystalline samples
  of {Y2O3}:{Eu3}+,'' {\em Physical Review B}, vol.~49, no.~9, p.~5821, 1994.

\bibitem{kozankiewicz_single-molecule_2014}
B.~Kozankiewicz and M.~Orrit, ``Single-molecule photophysics, from cryogenic to
  ambient conditions,'' {\em Chem. Soc. Rev.}, vol.~43, no.~4, pp.~1029--1043,
  2014.

\bibitem{serrano_all-optical_2018}
D.~Serrano, J.~Karlsson, A.~Fossati, A.~Ferrier, and P.~Goldner, ``All-optical
  control of long-lived nuclear spins in rare-earth doped nanoparticles,'' {\em
  Nature Communications}, vol.~9, p.~2127, May 2018.

\bibitem{afzelius_impedance-matched_2010}
M.~Afzelius and C.~Simon, ``Impedance-matched cavity quantum memory,'' {\em
  Physical Review A}, vol.~82, no.~2, p.~022310, 2010.

\bibitem{macfarlane_optical_2007}
R.~M. Macfarlane, ``Optical {Stark} spectroscopy of solids,'' {\em Journal of
  Luminescence}, vol.~125, no.~1-2, pp.~156--174, 2007.

\bibitem{altner_dephasing-rephasing_1996}
S.~B. Altner, M.~Mitsunaga, G.~Zumofen, and U.~P. Wild, ``Dephasing-{Rephasing}
  {Balancing} in {Photon} {Echoes} by {Excitation} {Induced} {Frequency}
  {Shifts},'' {\em Physical Review Letters}, vol.~76, pp.~1747--1750, Mar.
  1996.

\bibitem{wernsdorfer_synthetic_2019}
W.~Wernsdorfer and M.~Ruben, ``Synthetic {Hilbert} {Space} {Engineering} of
  {Molecular} {Qudits}: {Isotopologue} {Chemistry},'' {\em Advanced Materials},
  vol.~31, p.~1806687, June 2019.

\bibitem{godfrin_operating_2017}
C.~Godfrin, A.~Ferhat, R.~Ballou, S.~Klyatskaya, M.~Ruben, W.~Wernsdorfer, and
  F.~Balestro, ``Operating {Quantum} {States} in {Single} {Magnetic}
  {Molecules}: {Implementation} of {Grover}’s {Quantum} {Algorithm},'' {\em
  Physical Review Letters}, vol.~119, p.~187702, Nov. 2017.

\bibitem{kindem_control_2020}
J.~M. Kindem, A.~Ruskuc, J.~G. Bartholomew, J.~Rochman, Y.~Q. Huan, and
  A.~Faraon, ``Control and single-shot readout of an ion embedded in a
  nanophotonic cavity,'' {\em Nature}, vol.~580, pp.~1--12, Mar. 2020.

\end{thebibliography}

\section*{Acknowledgments}
We thank Mikael Afzelius for useful discussions.
This project has received funding from the European Union?s Horizon 2020 research and innovation programme under grant agreements No 820391 (SQUARE) and the ANR UltraNanoSpec projects, grant ANR-20-CE09-0022 of the French Agence Nationale de la Recherche.

\section*{Authors contributions}
P. G., M. R. and D. H. conceived and supervised the project. D. S. and K. S. K. were involved in the conceptual development of the project. 
K.S.K. and M.R. were responsible for the synthesis and characterisation of the isotopologue complexes. B. H. performed powder X-ray diffraction studies and indexed the patterns. O. F. solved the X-ray structure of the complex. D. S. and P. G. performed the optical experiments and analyzed the results. 
D. S. and P. G. wrote the manuscript with inputs form all authors.

\end{document}